\documentclass[%
 reprint,
 amsmath,amssymb,
 aps,
 prl,
 floatfix,
]{revtex4-2}
\usepackage{nicematrix}
\usepackage{mathtools}
\usepackage{graphicx}
\usepackage{dcolumn}
\usepackage{bm}

\usepackage[utf8]{inputenc} 
\usepackage{url}
\usepackage{amsmath}
\usepackage{amssymb}    
\usepackage{graphicx}
\usepackage{listings}
\usepackage{hyperref}

\renewcommand{\v}[1]{\ensuremath{\mathbf{#1}}} 
\newcommand{\gv}[1]{\ensuremath{\mbox{\boldmath$ #1 $}}} 
\newcommand{\grad}[1]{\nabla #1} 
\renewcommand{\div}[1]{\nabla \cdot #1} 
\newcommand{\pd}[2]{\frac{\partial #1}{\partial #2}} 
 
\renewcommand{\d}[2]{\frac{d #1}{d #2}} 
\newcommand{\delpar}[1]{\nabla_{\parallel} #1} 
\newcommand{\gradperp}[1]{\grad{}_{\perp} {#1}} 

\newcommand{\curv}[1]{{C}_{\left({#1}\right)}}

\newcommand{\cs}{c_s}

\newcommand{\bhatZ}{\v{b_0}}

\newcommand{\cur}{j_{\parallel}}

\newcommand{\vpi}{v_{\parallel i}}						
\newcommand{\vpe}{v_{\parallel e}}					
\newcommand{\gvort}{\omega}		

\newcommand{\momSrce}{S_{\mathcal{M}\parallel e}}							
\newcommand{\momSrci}{S_{\mathcal{M}\parallel i}}							
\newcommand{\nSrcN}{S_n}									
\newcommand{\enerSrceN}{S_{E,e}}							
\newcommand{\enerSrciN}{S_{E,i}}							

\newenvironment{eqnal}{\equation\aligned}{\endaligned\endequation}

\newcommand{\fortranl}[1]{\lstinline[language={[90]Fortran}]{#1}}

\begin{document}

\preprint{APS/123-QED}
\title{Deep electric field predictions by drift-reduced Braginskii theory with plasma-\\neutral interactions based upon experimental images of boundary turbulence}

\author{A. Mathews$^1$} \altaffiliation[Present address: ]{Swiss Plasma Center,  \'Ecole Polytechnique F\'ed\'erale de Lausanne, Lausanne, Vaud 1015 CHE} \email{mathewsabhilash@gmail.com} 
\author{J.W. Hughes$^1$} \author{J.L. Terry$^1$} \author{S.G. Baek$^1$}
\affiliation{%
$^1$Plasma Science and Fusion Center, Massachusetts Institute of Technology, Cambridge, Massachusetts 02139, USA
}%

\date{\today}

\begin{abstract}

We present 2-dimensional turbulent electric field calculations via physics-informed deep learning consistent with (i) drift-reduced Braginskii theory under the framework of an axisymmetric fusion plasma with purely toroidal field and (ii) experimental estimates of the fluctuating electron density and temperature on open field lines obtained from analysis of gas puff imaging of a discharge on the Alcator C-Mod tokamak. The inclusion of effects from the locally puffed atomic helium on particle and energy sources within the reduced plasma turbulence model are found to strengthen correlations between the electric field and electron pressure. The neutrals are also directly associated with broadening the distribution of turbulent field amplitudes and increasing ${\bf E \times B}$ shearing rates. This demonstrates a novel approach in plasma experiments by solving for nonlinear dynamics consistent with partial differential equations and data without encoding explicit boundary nor initial conditions.

\end{abstract}

\maketitle


Reduced turbulence models are, by definition, simplified descriptions of chaotic physical systems. Arguments for scale separation and geometric approximations are often undertaken in the pursuit of expedient yet reasonable estimates, but their precise effects on nonlinear processes in turbulence calculations are not always fully understood. As boundary plasmas in magnetic confinement fusion are governed by a vast range of spatiotemporal dynamics, model approximations are inevitable even using modern computing, but may only be weakly valid (if at all). To unequivocally quantify such impacts in nonlinear calculations, this Letter examines the fundamental connection between the turbulent fields predicted by a reduced model for a tokamak plasma by integrating the physics-informed deep learning technique from \cite{Mathews2021PRE} with experiment. Namely, based upon electron pressure measurements inferred from experimental images \cite{mathews2022deep}, we compute the 2-dimensional turbulent electric field consistent with electrostatic drift-reduced Braginskii fluid theory under the assumption of axisymmetry with a purely toroidal field. Previous work in modelling low-$\beta$ plasmas found excellent agreement when comparing the two-fluid theory's turbulent electric field to electromagnetic gyrokinetics \cite{Mathews2021PoP}. As an important first step towards translating the technique to experiment and directly testing reduced turbulence models, this Letter explores an edge plasma in the Alcator C-Mod tokamak \cite{Hutch_CMod,Marmar_CMod,Alcator_Greenwald}. All neglected physics in the applied turbulence theory can be re-inserted to ascertain their individual impacts, and as an initial probe, we test the inclusion of helium gas (which is locally puffed in the experiment) to gauge perturbative effects of injected neutral atoms, e.g. via the gas puff imaging (GPI) diagnostic \cite{Zweben_2017}. Past simulations \cite{Thrysoe_2018,Zholobenko_2021_validation} and experiments \cite{exp0neut,exp1neut,exp2neut} have investigated the role of neutrals on edge turbulent fields, although the applied methods (e.g. boundary conditions, biasing) are questionable and/or lacking the required precision for clearly testing nonlinear dynamical relationships. Conventional numerical models are challenged by the immense difficulty in aligning (1) initializations of fields, (2) boundary conditions, and (3) physical equations with chaotic systems. By coupling the independent techniques from \cite{Mathews2021PRE} and \cite{mathews2022deep}, we present a first-of-its-kind turbulence analysis in a nuclear fusion experiment which can bypass (1) and (2) to strictly focus on (3). In this Letter, our deep learning calculation finds that density and energy sources associated with time-dependent ionization of helium directly cause broadening in the turbulent electric field amplitudes along with an enhancement in correlation with electron pressure that is not otherwise extant in plasmas without such neutral processes. These effects, which are due to plasma-neutral interactions, reveal stronger ${\bf E \times B}$ flows and elevated average shearing rates on turbulent scales than expected in fully ionized gases.

The focus of our present analysis will be plasma discharge 1120711021 from Alcator C-Mod. This lower single null diverted ohmic L-mode plasma has an on-axis toroidal magnetic field of 5.4 T and plasma current of 0.8 MA. Based on the Thomson scattering and electron cyclotron emission diagnostics \cite{Basse_CMod}, the core electron density and temperature are $2.0 \times 10^{20} \ \text{m}^{-3}$ and $1.5$ keV, respectively. At the last closed flux surface (LCFS), they are roughly $3.0 \times 10^{19} \ \text{m}^{-3}$ and $40$ eV, respectively \cite{mathews2022deep}. Recent advancements \cite{Mathews2021PRE} permit the prediction of turbulent fields directly consistent with theory in experimental fusion conditions such as this tokamak plasma using helium GPI measurements \cite{GPImanual,mathews2022deep}. For this task, neural networks are applied to represent the drift-reduced Braginskii equations in the electrostatic limit relevant to low-$\beta$ conditions \cite{Mathews2021PoP}. Our turbulent electric field calculation assumes the 2-dimensional experimental inferences of the electron density and temperature are field-aligned \cite{Mathews2021PRE}, but since the GPI system on Alcator C-Mod views the edge plasma in the $(R,Z)$-plane \cite{mathews2022deep}, this results in the approximation of a purely toroidal magnetic geometry. The plasma is further assumed to be magnetized, collisional, and quasineutral with the perpendicular fluid velocity given by ${\bf E \times B}$, diamagnetic, and ion polarization drifts. After neglecting collisional drifts, as well as terms of order $m_e/m_i$, one arrives at the following set of equations (in Gaussian units) governing evolution of the plasma theory's density ($n \approx n_e$), vorticity ($\gvort$), parallel electron velocity ($\vpe$), parallel ion velocity ($\vpi$), electron temperature ($T_e$), and ion temperature ($T_i$) \cite{francisquez2020fluid}:
\begin{eqnal}\label{eq:nDotGDBH}
\d{^e n}{t} &= -\frac{2c}{B}\left[n\curv{\phi}-\frac{1}{e}\curv{p_e}\right]-n\delpar{\vpe} +\nSrcN+\mathcal{D}_{n},
\end{eqnal}
\begin{eqnal}\label{eq:wDotGDBH}
\pd{\gvort}{t} &= \frac{2c}{eB}\left[\curv{p_e}+\curv{p_i}\right]-\frac{1}{em_i \Omega_i}\curv{G_i} \\
&\quad+\frac{1}{e}\delpar{\cur}-\div{\left\lbrace\frac{nc^2}{\Omega_i B^2}\left[\phi,\gradperp{\phi}+\frac{\gradperp{p_i}}{en}\right]\right. \\ 
&\quad\left.+\frac{nc}{\Omega_i B}\vpi\delpar{\left(\gradperp{\phi}+\frac{\gradperp{ p_i}}{en}\right)}\right\rbrace}+\mathcal{D}_{\gvort},
\end{eqnal}
\begin{eqnal}\label{eq:vpeDotGDBH}
\d{^e\vpe}{t} &= \frac{1}{m_e}\left(e\delpar{\phi}-\frac{\delpar{p_e}}{n}-0.71\delpar{T_e} + e\eta_\parallel\cur \right) \\
&\quad + \frac{2}{3} \frac{\delpar{G_e}}{n} + \frac{2cT_e}{eB}\curv{\vpe}+\momSrce+\mathcal{D}_{\vpe},
\end{eqnal}
\begin{eqnal}\label{eq:vpiDotGDBH}
\d{^i\vpi}{t} &= \frac{1}{m_i}\left(-e\delpar{\phi}-\frac{\delpar{p_i}}{n}+0.71\delpar{T_e} - e\eta_\parallel\cur \right)\\
&+\frac{2}{3}\frac{\delpar{G_i}}{n}-\frac{2cT_i}{eB}\curv{\vpi}+\momSrci+\mathcal{D}_{\vpi},
\end{eqnal}
\begin{eqnal}\label{eq:TeDotGDBH}
\d{^e T_e}{t} = \frac{2T_e}{3n}\left[\d{^e n}{t} + \frac{1}{T_e}\delpar \kappa^e_\parallel \delpar T_e + \frac{5n}{m_e \Omega_e} \curv{T_e} \right.\\ \left. + \eta_\parallel \frac{\cur^2}{T_e} + \frac{0.71}{e}(\delpar{\cur} - \frac{\cur}{T_e}\delpar{T_e}) + \frac{1}{T_e} \enerSrceN \right] + \mathcal{D}_{T_e},
\end{eqnal}
\begin{eqnal}\label{eq:TiDotGDBH}
\d{^i T_i}{t} &= \frac{2T_i}{3n}\left[\d{^i n}{t} + \frac{1}{T_i}\delpar \kappa^i_\parallel \delpar T_i \right.\\
&\quad \left. - \frac{5n}{m_i \Omega_i} \curv{T_i} + \frac{1}{T_i} \enerSrciN \right] + \mathcal{D}_{T_i},
\end{eqnal}
whereby the field-aligned electric current density is $\cur = en\left(\vpi - \vpe\right)$, the stress tensor's gyroviscous terms contain $G_s = \eta^s_0 \left\lbrace 2\delpar{v_{\parallel s}}+c\left[\curv{\phi} + \curv{p_s}/(q_s n)\right]\right\rbrace$, and $\eta^s_0$, $\Omega_s$, and $q_s$ are the species ($s = \{e,i\}$) viscosity, cyclotron frequency, and electric charge, respectively. The convective derivatives are $d^s f/dt = \partial_t f + (c/B)\left[\phi,f\right] + v_{\parallel s}\delpar{f}$ with $\left[F,G\right] = \bhatZ \times \nabla F \cdot \nabla G$ and $\bhatZ$ representing the unit vector parallel to the background magnetic field. Consistent with Alcator C-Mod, the minor and major radius are $a_0 = 0.22 \text{ m}$ and $R_0 = 0.68 \text{ m}$, respectively, and there is a $1/R$ variation in the magnetic field strength, $B$, arising from the toroidal field coils. This gives rise to a curvature of $\gv{\kappa} = -{\bf{\hat{R}}}/R$, and curvature operator given by $\curv{f} = \bhatZ \times \gv{\kappa} \cdot \grad{f}$. The coefficients $\kappa^s_\parallel$ and $\eta^s_\parallel$ correspond to parallel heat diffusivity and conductivity, respectively. Ordinarily, the electrostatic potential, $\phi$, is computed via solving the following boundary value problem

\begin{equation}
\div{ \frac{nc}{\Omega_i B}\left(\gradperp{\phi}+\frac{\gradperp{p_i}}{en}\right) } = \gvort,
\end{equation}

\noindent but we instead follow the prescription developed in \cite{Mathews2021PRE} which utilizes just field-aligned turbulent electron pressure measurements along with Eqs. \eqref{eq:nDotGDBH} and \eqref{eq:TeDotGDBH} to calculate $\phi$.
No boundary nor initial conditions are explicitly assumed within our physics-informed deep learning framework. All analytic terms encoded in these continuum equations are computed exactly by the graph networks without any approximation as this machine learning framework uses a continuous spatiotemporal domain (e.g. no linearization nor discretization). Hyperdiffusion ($\mathcal{D}$), which is ordinarily applied for stability in numerical codes, is set to zero. Density ($\nSrcN$) and energy ($S_{E,s}$) sources associated with time-dependent ionization of the local helium gas based upon collisional radiative modelling are outlined in \cite{mathews2022deep}. In particular, we set $\nSrcN = n_0 n_e S_{CR}$ and $S_{E,e} = -E_{HeI} \nSrcN$, where $n_0$ is the atomic helium density, $S_{CR}$ corresponds to the ionization rate coefficient, and $E_{HeI} = 24.587$ eV is the ionization energy of HeI. The 2-dimensional turbulent $n_e$ and $T_e$ in experiment come from an $(R,Z)$-aligned plane on open field lines with a rectangular cross-section that spans a region of the scrape-off layer over $[90.3 < R \ \text{(cm)} < 90.9, -4.6 < Z \ \text{(cm)} < -1.0]$ for a duration of $1.312799 < t_{GPI} \ \text{(s)} < 1.312896$. The spatial and temporal resolution for the GPI diagnostic are approximately 1--2 mm and 2.5 $\mu$s, respectively. By assuming $\bhatZ$ to be parallel to these measurements, our plasma turbulence model essentially neglects the poloidal component of the field lines present in Alcator C-Mod. For physical orientation, the local magnetic field lines point towards the imaging system and $\bf{B} \times \nabla B$ is directed downwards. Moreover, in keeping with the axisymmetric approximation in \cite{Mathews2021PRE,Mathews2021PoP}, gradients along this field-aligned (nominally toroidal) direction are assumed to be small, i.e. $\nabla_\parallel \rightarrow 0$. Accordingly, an orthogonal right-handed geometry is employed for modelling whereby $x \equiv R$ is the radial coordinate, the parallel coordinate $\bhatZ$ is purely toroidal, and the binormal (nominally vertical) direction is $y \equiv Z$. The plasma theory consists of electrons and deuterium ions with real electron-ion mass ratio, i.e. $m_i = 3.34 \times 10^{-27} \text{ kg}$ and $m_e = 9.11\times 10^{-31} \text{ kg}$. Beyond the inclusion of appropriate sources and collisional drifts, our validated technique \cite{Mathews2021PRE} to calculate the turbulent electric field is applicable even if multiple ions and impurities are present in the experimental plasma due to quasi-neutrality underlying the electron fluid theory \cite{multi_species}.

\begin{figure*}[ht]
\includegraphics[width=1.0\linewidth]{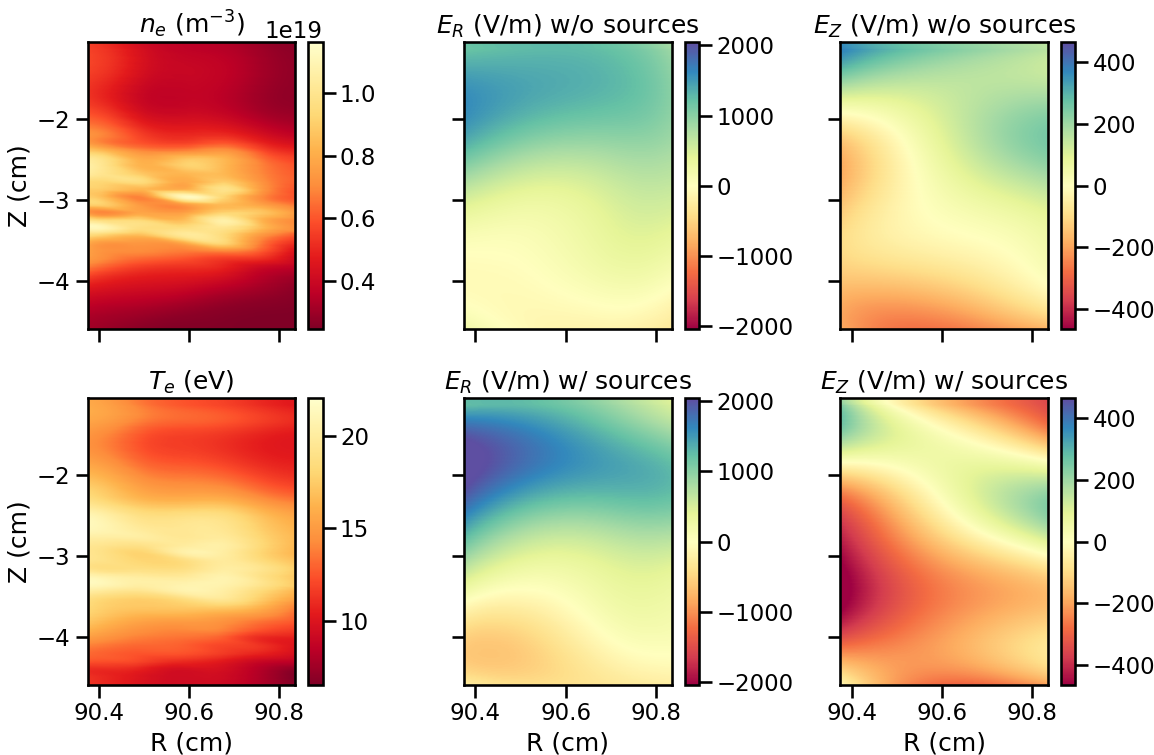}
\caption{\label{observed_neTe_predicted_Er}The 2-dimensional $n_e$ and $T_e$ (leftmost column) come from experimental GPI of discharge 1120711021 on Alcator C-Mod at $t = 1.312886$ s \cite{mathews2022deep}. The domains analyzed on these open field lines roughly span extents of 0.6 cm and 3.6 cm in the $R-$ and $Z-$directions, respectively. The $E_R$ and $E_Z$ are inferred from drift-reduced Braginskii theory using these experimental $n_e$ and $T_e$ according to the outlined deep learning framework \cite{Mathews2021PRE} in the limiting cases of with (i.e. scaling factor of $n_0^* = 10^{19}$ m$^{-3}$) and without (i.e. $n_0^* = 0$) HeI sources in the physical equations.}
\end{figure*}

\begin{table}[ht]
\centering
\renewcommand{\arraystretch}{1.}
\includegraphics[width=1.0\linewidth]{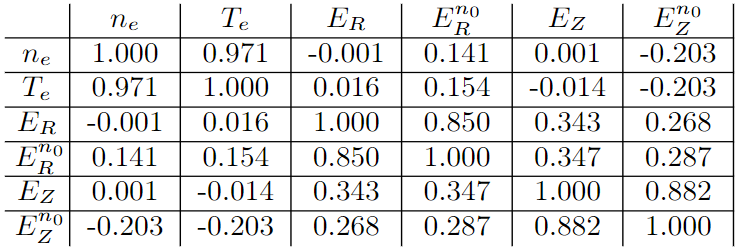}
\caption{\label{table_correlation_matrix}A correlation matrix of the turbulent fluctuations where $n_e$ and $T_e$ are inferred from plasma discharge 1120711021 based upon experimental GPI measurements roughly over $90.3 < R \ \text{(cm)} < 90.9$, $-4.6 < Z \ \text{(cm)} < -1.0$, and $1.312799 < t_{GPI} \ \text{(s)} < 1.312896$. The quantities $E_R^{n_0}$ and $E_R$ ($E_Z^{n_0}$ and $E_Z$) in this table correspond to the radial (vertical) turbulent electric fields predicted by drift-reduced Braginskii theory with (i.e. scaling $n_0^*$ to $10^{19}$ m$^{-3}$) and without (i.e. setting $n_0^*$ to $0$) HeI sources, respectively.}
\end{table}

As visualized in Figure \ref{observed_neTe_predicted_Er}, using the 2-dimensional $(R,Z)$-aligned experimentally-inferred $n_e$ and $T_e$ measurements from the helium GPI diagnostic, we compute the turbulent electric field predicted by the ascribed drift-reduced Braginskii theory in the limits of (i) no sources and (ii) source effects due to time-dependent ionization of HeI. Since only the relative (and not absolute) brightness of the line emission across the field-of-view of the GPI is known for this plasma discharge, only the structure of the experimental $n_0$ can be inferred. Nevertheless, as a conservative lower bound on $n_0$ in the calculations to test the impacts of neutral dynamics on turbulent fields, the atomic helium density is scaled to an amplitude of approximately $10^{19}$ m$^{-3}$ \cite{SGBAEK_DEGAS2_GPI_CMOD}. Much larger scaling factors (e.g. $n_0^* = 10^{20}$ m$^{-3}$) were found to lead to numerical instability in the optimization, which suggests mathematical terms (e.g. poloidal flows) are missing in the model equations and/or that such high $n_0$ are unphysical. A matrix of correlation coefficients for these fluctuations is given in Table \ref{table_correlation_matrix}. The correlations between the turbulent electric field and electron pressure predicted by theory in fully ionized conditions are found to be nearly zero. This nonlinear connection changes with the inclusion of plasma-neutral interactions: time-dependent ionization effects due to atomic helium induce a positive (negative) dependence for $E_R$ ($E_Z$) on $n_e$ and $T_e$. If the experimental $n_0$ is truly larger in magnitude, the reported correlations in Table \ref{table_correlation_matrix} are expected to be even stronger.

\begin{figure}
\includegraphics[width=1.0\linewidth]{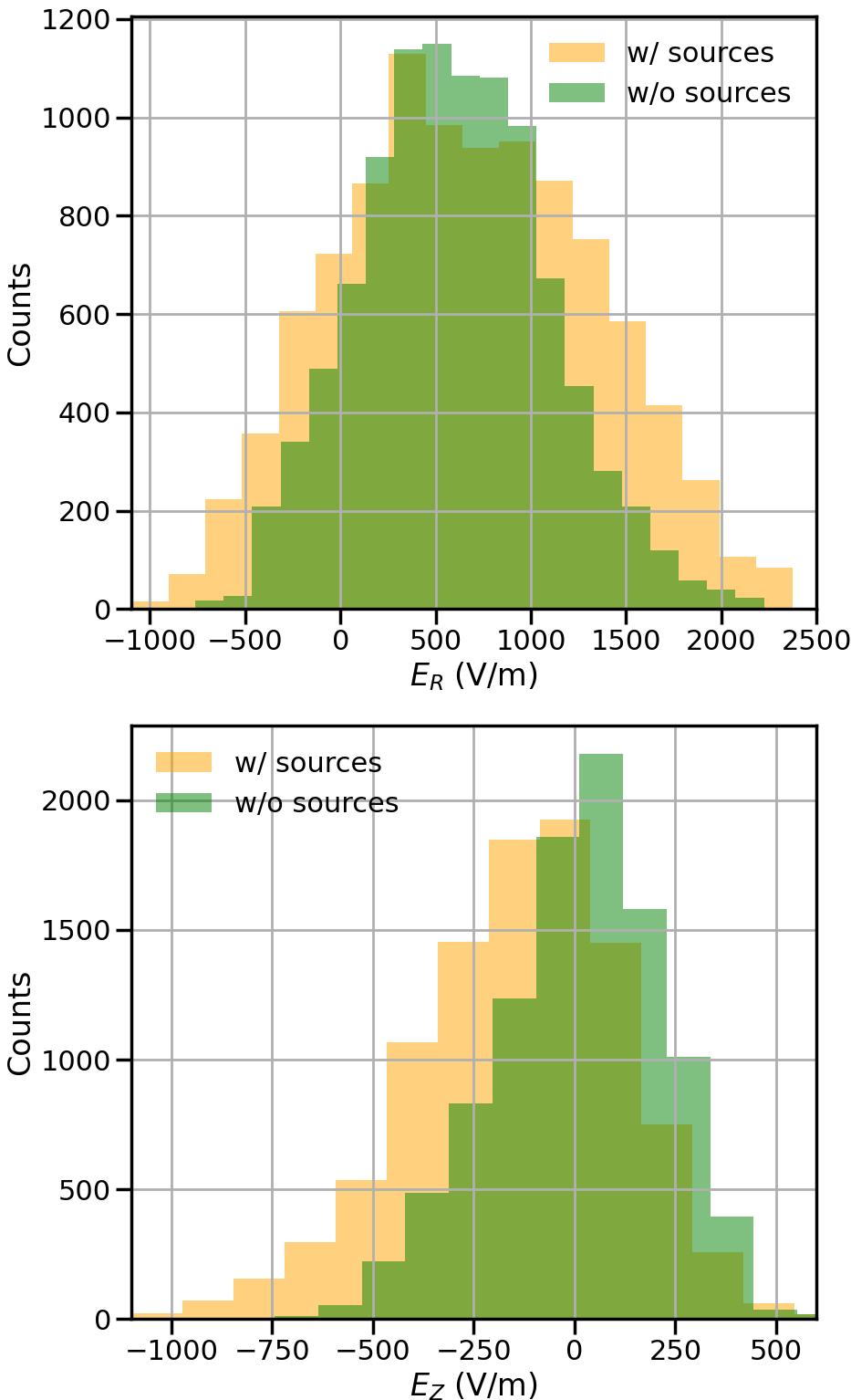}
\caption{\label{histogram_Er}Histograms of $E_R$ and $E_Z$ consistent with drift-reduced Braginskii theory in a toroidal axisymmetric geometry evaluated at the GPI pixels from $90.3 < R \ \text{(cm)} < 90.9$, $-4.6 < Z \ \text{(cm)} < -1.0$, and $1.312799 < t_{GPI} \ \text{(s)} < 1.312896$ in discharge 1120711021 on Alcator C-Mod.}
\end{figure} 

\begin{figure*}[ht]
\includegraphics[width=1.0\linewidth]{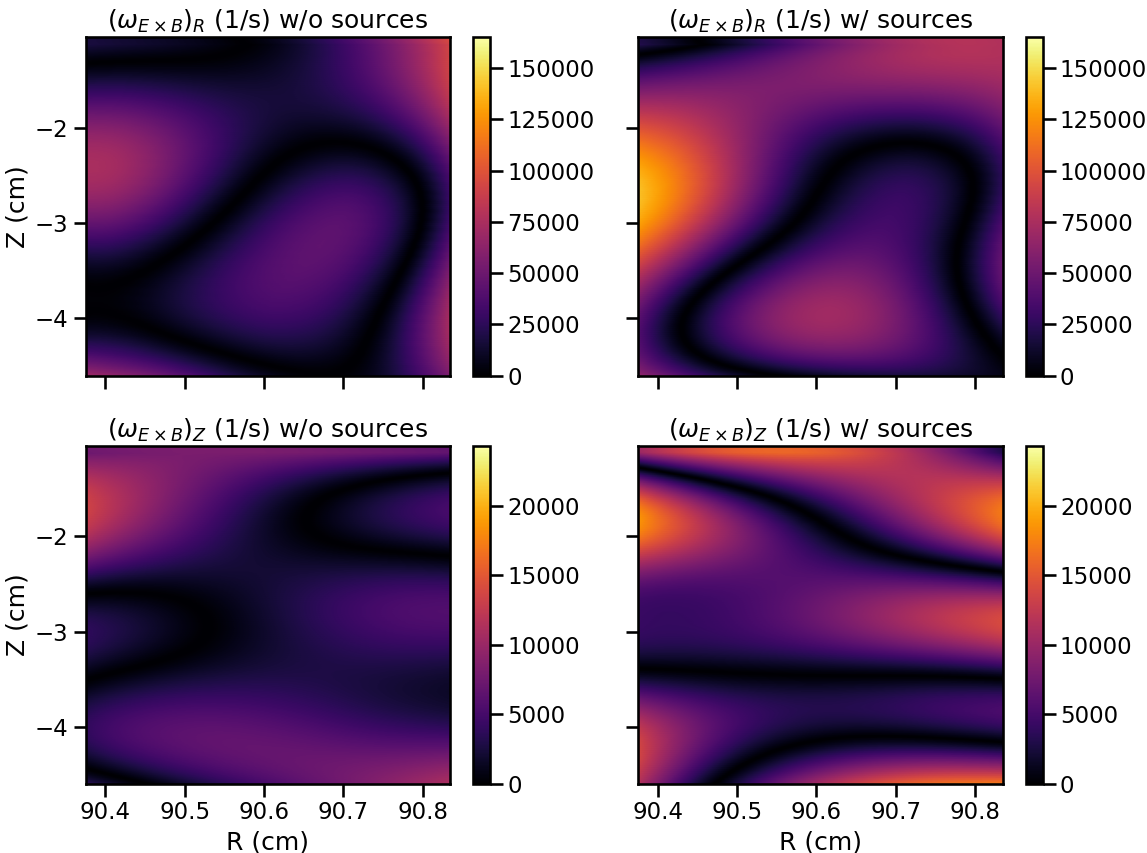}
\caption{\label{ExB_velocity_shear}Visualizations of the radial and vertical turbulence shearing rates predicted by drift-reduced Braginskii theory in discharge 1120711021 at $t = 1.312886$ s on Alcator C-Mod under the assumption of axisymmetry with a toroidal field. The plots consider no sources (left) and, alternatively, neutral sources to account for time-dependent ionization of HeI (right).}
\end{figure*}

Further, the addition of neutral helium to drift-reduced Braginskii theory are found to broaden the distribution of turbulent field magnitudes over the 2-dimensional domain as displayed in Figure \ref{histogram_Er}. This leads to amplified electric field fluctuations with sharper radial variation in the electric potential structure, and manifests as larger ${\bf E \times B}$ flows on turbulent scales in the boundary plasma. Intuitively, this arises since the observed spatiotemporal evolution of $n_e$ and $T_e$ is not solely due to transport, but instead the self-consistent turbulent ${\bf E \times B}$ flows have to be mathematically balanced in Eqs. \eqref{eq:nDotGDBH} and \eqref{eq:TeDotGDBH} with sources. Experimentally, such effects are important for turbulence spreading \cite{Grenfell_2018} and material interactions since even small drifts can compete with flows perpendicular to surfaces at the plasma-sheath interface \cite{edge_flows}. Additionally, the radial and vertical turbulence shearing rates, $({\omega_{\bf E \times B}})_R = \lvert \partial({v_{\bf E \times B}})_Z/\partial R \rvert$ and $({\omega_{\bf E \times B}})_Z = \lvert \partial({v_{\bf E \times B}})_R/\partial Z \rvert$, are elevated on average when atomic helium is present in the edge compared to the case without time-dependent ionization. At intermediate $n_0$, $\langle ({\omega_{\bf E \times B}})_R \rangle$ and $\langle ({\omega_{\bf E \times B}})_Z \rangle$ still increase with $n_0$, although the trend is not strictly linear, as displayed in Table \ref{table_shear}. The modified $\omega_{\bf E \times B}$ visualized in Figure \ref{ExB_velocity_shear} can impact shear flow stabilization and cross-field transport of coherent structures. Not including time-dependent neutrals in nonlinear simulations can accordingly mask these effects. All in all, the amplification of fields due to atomic helium in a deuterium plasma and presence of correlations not otherwise present in fully ionized gases quantitatively demonstrate perturbative effects on turbulent scales from diagnostics which locally inject neutrals. They should thus be accounted in experimental tests to precisely validate edge turbulence models, otherwise such errors in fields due to plasma-neutral interactions that scale nonlinearly with $n_0$ will exist.



\begin{table}[ht]
\centering
\renewcommand{\arraystretch}{1.}
\includegraphics[width=0.85\linewidth]{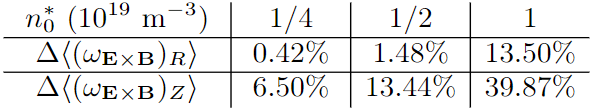}
\caption{\label{table_shear}Change in nonlinear turbulence shearing rates computed at varying HeI densities and averaged over the spatiotemporal domain spanned by the camera frames. These calculations are relative to the case with no sources where $\langle ({\omega_{\bf E \times B}})_R \rangle = 4.74 \times 10^4$ s$^{-1}$ and $\langle ({\omega_{\bf E \times B}})_Z \rangle = 9.08 \times 10^3$ s$^{-1}$.}
\end{table}

Going forward, we point out that the time-dependent 2-dimensional $n_e$ and $T_e$ are based upon generalized collisional radiative constraints that are agnostic to any turbulence model. This permits the first ever self-consistent learning of time-dependent 2-dimensional profiles for neutral species such as atomic and molecular deuterium \cite{coroado2021selfconsistent} via application of existing Monte Carlo transport codes \cite{STOTLER_time_dependent_DEGAS2}, which could be playing a significant role---as exemplified above by ionization of atomic helium---and can be added into our computational framework for model validation efforts with fuller fidelity to experiment. By isolating these effects such as the broadening of electric field amplitudes and shearing rates due to atomic helium, we can quantitatively identify essential physics in the development of effective reduced turbulence models. Overall, this Letter illustrates a novel pathway towards uncovering unobserved dynamics in experimental fusion plasmas which are conventionally difficult to diagnose. Further, by making no explicit assumptions on boundary conditions for turbulent fields within the physics-informed deep learning framework, we are now able to precisely test the nonlinear impacts of critical approximations (e.g. neglecting time-dependent neutral sources) on plasma dynamics in these chaotic systems.

\begin{acknowledgments}
 We wish to thank  M. Francisquez for insights shared and helpful discussions. All codes were run using MIT's Engaging cluster and we are grateful for the team's assistance with computing resources. The work is supported by the Natural Sciences and Engineering Research Council of Canada (NSERC) by the doctoral postgraduate scholarship (PGS D), Manson Benedict Fellowship, and the U.S. Department of Energy (DOE) Office of Science under the Fusion Energy Sciences program by contracts DE-SC0014264 and DE-SC0014251. Relevant data and files are available from the corresponding author.
\end{acknowledgments}

\bibliography{main.bib}
\bibliographystyle{apsrev4-2}
\end{document}
%